\documentclass[%
 reprint,
 amsmath,amssymb,
 prl,
]{revtex4-2}

\usepackage{graphicx}
\usepackage{dcolumn}
\usepackage{bm}
\usepackage{color}
\usepackage{float}

\usepackage{lipsum}%

\begin{document}


\title[Scaling law in electrified droplets]{Universal scaling law for electrified sessile droplets on a lyophilic surface}

\author{Dipin S. Pillai$^1$\footnote{dipinsp@iitk.ac.in} and Kirti Chandra Sahu$^2$\footnote{ksahu@che.iith.ac.in}}
\affiliation{ 
$^1$Department of Chemical Engineering, Indian Institute of Technology Kanpur, Uttar Pradesh, 208016, India \\
$^2$Department of Chemical Engineering, Indian Institute of Technology Hyderabad, Sangareddy, Telangana, 502 284, India
}%

\date{\today}

\begin{abstract}
Electrified sessile droplets on solid surfaces are ubiquitous in nature as well as several practical applications. Although the influence of electric field on pinned sessile droplets and soap bubbles have been investigated experimentally, the theoretical understanding of the stability limit of generic droplets remains largely elusive. By conducting a theoretical analysis in the framework of lubrication approximation, we show that the stability limit of a sessile droplet on a lyophilic substrate in the presence of an electric field exhibits a universal power-law scaling behaviour. The power-law exponent between the critical electric field and the droplet volume is found to be -1. The existence of this scaling law is further explained by virtue of minimization of the total free energy of the electrified droplet.
\end{abstract}

\keywords{Suggested keywords}
\maketitle


Due to its importance in a number of applications, such as meteorology \cite{rayleigh1882xx,sartor1969electricity}, nuclear research \cite{bohr1939mechanism}, biological applications \cite{collins2008electrohydrodynamic}, microfluidics, and electrospraying \cite{collins2013universal}, droplet actuation under electric fields has been investigated since the early 1900s. The interface between two fluids having distinct electrical and hydrodynamic properties, such as the one between a droplet and its ambient medium, deforms along the direction of the applied electric field and undergoes fragmentation when the applied electric field exceeds a critical limit \cite{duft2003rayleigh,vlahovska2019electrohydrodynamics}. \citet{rayleigh1882xx} and \citet{wilson1925bursting} were the first to use analytical and experimental methods respectively to study the stability of electrified droplets. \citet{wilson1925bursting} demonstrated that increasing the applied electric field causes a half-sphere soap bubble to elongate into a steady egg shape (referred to here as a stable droplet), which upon further increase in electric field results in a jet-like structure emanating from the droplet apex (referred to here as an unstable droplet). The disintegration of the electrified free-floating droplets and the resultant tip streaming along the direction of electric field have been shown to follow a universal scaling law between the droplet charge and its radius \cite{collins2013universal}. Determining the critical limit of the electric field below which a droplet can maintain its static shape is a challenging mathematical exercise. This is due to the non-trivial relationship of the shape of the droplet and the pressure inside it with the applied electric field, which requires obtaining solution of coupled partial differential equations. Recently, some theoretical studies have focused on the deformation and breakup of free-floating droplets under an external electric field, and reported the existence of scaling behaviour at the critical condition, which have been thoroughly reviewed by \citet{vlahovska2019electrohydrodynamics}. The complexity of determining the stability limit in the case of a sessile droplet is substantially higher due to the presence of contact-line dynamics.



\citet{beroz2019stability} recently established a remarkable power-law scaling for the stability limit of electrified pinned sessile droplets having a constant radius in terms of two dimensionless groups, namely $R^3/V$ and $\varepsilon {E}^2 R/\gamma$, derived by minimising the free energy of the system, viz., a combination of surface energy and electrostatic energy. Here, the droplet's volume is $V$, its surface tension with respect to the ambient is $\gamma$, the permittivity of the dielectric ambient medium is $\varepsilon$, and the applied uniform electric field is $E$. The droplet radius, $R$, appears as a characteristic length scale in their experiments since the droplet is constrained to have a constant radius, thereby limiting their analysis only to pinned droplets. However, the radius ($R$), contact angle ($\theta$) and the droplet's height $(h_c)$ are coupled, and not known a priori for a generic electrowetting scenario. As a result, $V^{1/3}$ is the only relevant length scale associated with droplet size. In addition, for our case, an additional non-dimensional number arises due to the length scale associated with electrode spacing, $H_0$. Thereby, the stability limit ought to be described by two dimensionless numbers, i.e., $\varepsilon {E}^2 H_0/\gamma$ and $H_0^3/ V$. 

\begin{figure}
\centering
\includegraphics[width=0.45\textwidth]{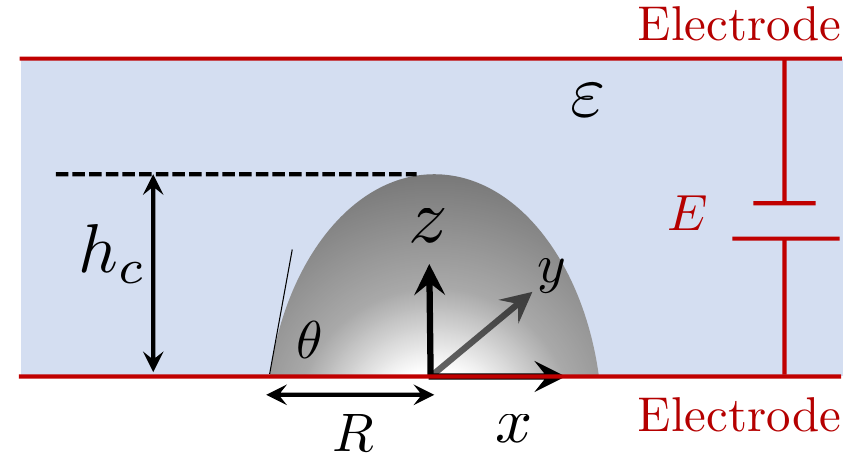} 
\caption{Schematic of a perfectly conducting sessile droplet with its centerline height, $h_c(t)$, under a constant electric field, $E$. The permittivity of the surrounding medium is $\varepsilon$ and $\gamma$ represents the surface tension of the droplet with respect to the surrounding medium.}
\label{fig:schematic}
\end{figure}

Unlike the constant wetting radius case of \citet{beroz2019stability}, we investigate a more generic scenario of electrowetting, wherein the wetting radius of the droplet is free to evolve in response to the applied electric field. This is indeed what we observe in reality. We consider a perfectly conducting droplet placed on a lyophilic substrate (where the static contact angle of the droplet, $\theta<90^\circ$), which acts as one of the electrodes, resulting in a potential difference of $\Phi (=EH_0)$ with respect to another electrode placed in a co-planar configuration as shown in figure \ref{fig:schematic}. An example of such a system is a water droplet placed on a glass substrate cleaned with piranha solution and dried under nitrogen atmosphere, which exhibits a contact angle of $12^\circ \pm 3^\circ$ \cite{mondal2018patterns}. Further, the water-air system is known to obey the perfect conductor model \cite{Ward2019}. At room temperature, water has a conductivity of 5.5 $\times 10^{-6}$ Sm$^{-1}$ \cite{SIdata}, which is several orders of magnitude greater than the conductivity of ambient air ($\sim 10^{-13}$ Sm$^{-1}$). Therefore, a perfect conductor-perfect dielectric model which is used in the current study ought to be valid for the water–air like systems \cite{Ward2019, pillai2018}. In our study, for a given droplet volume, we first determine the equilibrium droplet shape which is then used as an initial condition for the cases wherein electric field is applied. The physical parameters used in the numerical simulations are $\rho=1000$ kgm$^{-3}$, $\varepsilon=8.854\times10^{-12}$ Fm$^{-1}$,  $\mu=0.9\times10^{-3}$ Pa$\cdot$s and $\gamma=7.2\times10^{-2}$ Nm$^{-1}$.

\begin{figure*}
\centering
\includegraphics[width=0.9\textwidth]{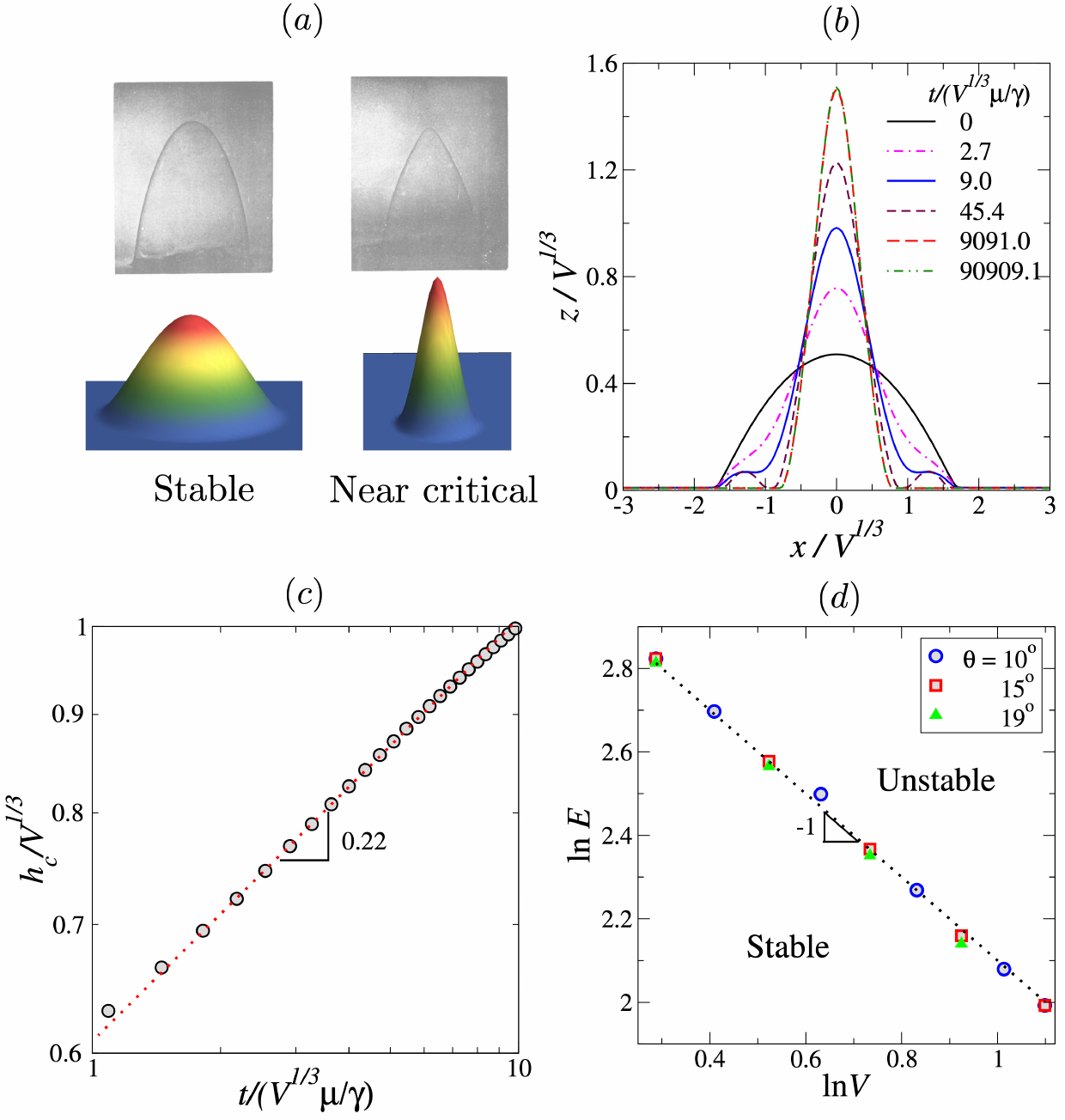} 
\caption{(a) Demonstration of stable and critically stable droplets with an increase in electric field. In this panel, the top row is from \citet{wilson1925bursting} and the bottom row depicts the droplet's shapes obtained from the present three-dimensional simulations. The stable and the critically stable droplets in our simulations correspond to $E=73.3$ kV/mm and 168.3 kV/mm, respectively, with the rest of the parameters $(\theta = 15^\circ$ and $V=1.33$ mm$^3$) remain the same in both the cases. (b) Temporal evolution of the droplet's shape at the critical electric field for $\theta = 15^\circ$, $E=168.3$ kV/mm and $V=1.33$ mm$^3$. (c) Variation of the normalised droplet's height $(h_c/V^{1/3})$ with normalised time $(t/(V^{1/3}\mu/\gamma)$ during its initial stage of deformation. At the initial stage, the deformation exhibits a power law with an exponent 0.22 for the case considered in panel (b). It is also observed that for the range of parameters considered in our study, the droplet exhibits the power-law behaviour, but the corresponding exponent depends on electric field and droplet's volume. (d) Variation of ${\rm ln}E$ (kV/mm) versus ${\rm ln} V$ (mm$^3$) showing a power law behaviour (exponent $\sim -1$) for $\theta =10^\circ$, $15^\circ$ and $19^\circ$. This scaling law is found to be universal in nature.}
\label{fig:fig2}
\end{figure*}

The three-dimensional evolution equations for a perfectly conducting droplet on a lyophilic substrate, obtained using the inertial lubrication model based on the method of weighted residual integral boundary layer theory, are given by \cite{pillai2021sessile,kainikkara2021equivalence}
\begin{subequations}\label{alleq}
\begin{eqnarray}
\frac{\partial h}{\partial t}+\frac{\partial q_x}{\partial x}+\frac{\partial q_y}{\partial y}=0, ~~~~~~~~~~~~~
\\
\int_0^h \rho F\Bigg( \frac{\partial {\hat u}}{\partial t}+ {\hat u} \frac{\partial {\hat u}}{\partial x} + {\hat v} \frac{\partial {\hat u}}{\partial y} + {\hat w} \frac{\partial {\hat u}}{\partial z} \Bigg)dz = \mu q_x  \nonumber \\ + \Bigg[\gamma\frac{\partial}{\partial x}\Bigg(\frac{\partial^2 h}{\partial {x}^2}+\frac{\partial^2 h}{\partial {y}^2}\Bigg) -\frac{\varepsilon\Phi^2}{(H_0-h)^3}\frac{\partial h}{\partial x} \nonumber \\ +\frac{2\gamma(1-\cos\theta)}{h_f} \frac{\partial}{\partial x}\Bigg(\frac{h_f^{3}}{h^3}-\frac{h_f^{2}}{h^2}\Bigg)\Bigg]\int_0^h F dz,
\\
\int_0^h \rho F\Bigg( \frac{\partial {\hat v}}{\partial t}+ {\hat u} \frac{\partial {\hat v}}{\partial x} + {\hat v}\frac{\partial {\hat v}}{\partial y} + \hat{w}\frac{\partial{\hat{v}}}{\partial{z}}\Bigg)dz = \mu q_y \nonumber \\ + \Bigg[\gamma\frac{\partial}{\partial y}\Bigg(\frac{\partial^2 h}{\partial {x}^2}+\frac{\partial^2 h}{\partial {y}^2}\Bigg)-\frac{\varepsilon\Phi^2}{(H_0-h)^3}\frac{\partial h}{\partial x}\nonumber \\
+\frac{2\gamma(1-\cos\theta)}{h_f} \frac{\partial}{\partial y}\Bigg(\frac{h_f^{3}}{h^3}-\frac{h_f^{2}}{h^2}\Bigg)\Bigg]\int_0^h F dz.
\end{eqnarray}
\end{subequations}
Here, $h$ is the interface height; $t$ is time; $(x,y,z)$ represent the Cartesian coordinate system as shown in figure \ref{fig:schematic}; $q_x$ and $q_y$ denote the depth-averaged flow rate in $x$ and $y$ directions, respectively; $\hat{u}$, $\hat{v}$ and $\hat{w}$ are the $\mathcal{O}(1)$ contributions of the velocity components in the $x$, $y$ and $z$ directions, respectively; $h_f$ represents the film thickness that minimizes the  conjoining-disjoining potential based on the Hamaker theory \cite{Gomba2010}; $\mu$ and $\rho$ are the dynamic viscosity and density of the droplet; $F$ is the Galerkin weight function. The governing equations are solved using the Fourier spectral collocation technique \cite{pillai2021sessile,kainikkara2021equivalence}. Although, we have used a Cartesian coordinate system, the problem can be identically formulated in the cylindrical polar coordinate system. Also, the current model is typically valid for Reynolds number of $\mathcal{O}(1)$, however, the stability limit is expected to be independent of the same. The detailed derivation of the governing equations, the associated boundary conditions and the solution technique can be found in the supplementary information.

In figure \ref{fig:fig2}(a), we show the typical stable and critically stable shapes of the droplet obtained by increasing the electric potential in our simulations along with the corresponding experimental results of \citet{wilson1925bursting}. Note that \citet{wilson1925bursting} used soap bubbles instead of liquid droplets and therefore only qualitative comparison is made. In our study, the critical stability limit was ascertained from transient simulations. For a given electric field, transient simulations were carried out until a steady droplet height was attained. Beyond a critical electric field, a steady solution was not obtained, and the interface was seen to sharply spike to rupture at the top wall. This critical field was used as the droplet's stability limit. In figure \ref{fig:fig2}(b), the normalized temporal evolution of a droplet of volume, $V=1.33$ mm$^3$ and $\theta = 15^\circ$ at the corresponding critical electric field, $E=168.3$ kV/mm, is depicted. It can be seen in this case that the droplet evolves into a prolate egg-like shape and subsequently elongates towards a spike-like structure near the apex. The droplet is unstable beyond this critical electric field. The initial transient dynamics of normalized droplet height ($h_c/V^{1/3}$) with normalized time  ($t/V^{1/3}\mu/\gamma$) exhibits a power-law with slope 0.22 as shown in figure \ref{fig:fig2}(c). We observed that, while the droplet exhibits a power-law for other set of parameters as well, the exponent of this power-law is dependent on $\theta$ and $V$. Figure \ref{fig:fig2}(d) demarcates the stable and unstable regions in ${\rm ln}E-{\rm ln} V$ parameter space for three different initial contact angles, namely $\theta=10^\circ$ (circle), $\theta=15^\circ$ (square) and $\theta=19^\circ$ (triangle). It is to be noted that the initial contact angle of the droplet corresponds to its equilibrium shape in the absence of the electric field. Under the influence of the electric field, the droplet elongates in the direction of the applied electric field, and thus its contact angle increases with time. The earlier reports have demonstrated the validity of lubrication model for contact angles up to $65^\circ$ \cite{charitatos2020thin,tembely2019comprehensive}. We observed that even for very small equilibrium contact angles considered in this study, the instantaneous contact angle increases considerably with time and saturates to a steady value for stable droplets. In case of unstable droplets, the instantaneous contact angle continues to increase until rupture. Thus, we restrict our analysis to the three initial contact angles of $\theta=10^\circ$, $15^\circ$ and $19^\circ$ in order to ensure that the droplet shape at its stability limit is well within the validity of lubrication theory. 

\begin{figure}
\centering
\includegraphics[width=0.45\textwidth]{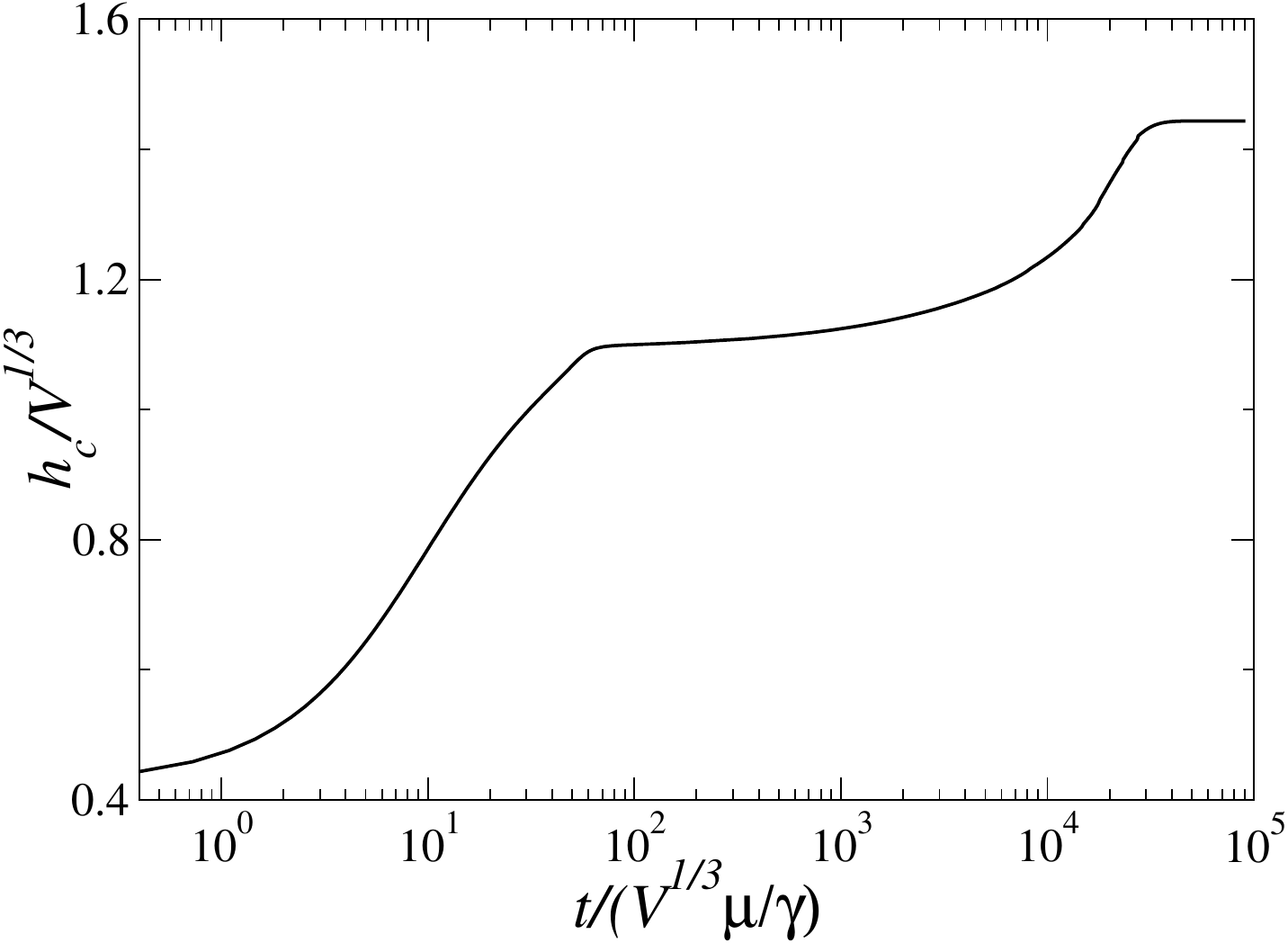} 
\caption{Temporal evolution of the normalised droplet height ($h_c/V^{1/3}$) for $\theta = 10^\circ$, $E=168.3$ kV/mm and $V=1.33$ mm$^3$.}
\label{fig:contact}
\end{figure}

It can be seen that the data in figure \ref{fig:fig2}(d) collapse to a single curve such that ${\rm ln}E_{cr} \sim {{\rm ln} V_{cr}}^{-1}$. Here, $E_{cr}$ is the critical limit of the electric field at which a droplet with volume $V_{cr}$ becomes unstable. Thus, this result demonstrates the existence of a universal power-law scaling for the critical field with droplet volume for the generic case of electrowetting on a lyophilic substrates. Figure \ref{fig:contact} demonstrates a typical long-time temporal evolution of the normalised droplet height ($h_c/V^{1/3}$) for a stable droplet. It can be observed that after the initial exponential transient as discussed in figure \ref{fig:fig2}(c), the droplet exhibits a secondary transient evolution, before finally settling at its steady configuration.

We now proceed to explain this universal power-law scaling behaviour using the minimization of the total free energy of the system. This idea is an extension of the one proposed by \citet{beroz2019stability}. As the wetting radius, $R$, is unknown and changes over time in the case of a generic droplet, we use $V^{1/3}$ as the length scale instead of $R$, which was employed by \citet{beroz2019stability} for pinned droplets. The total free energy, ${\cal F}$, of the system, is given by the sum of surface and electrical energies, i.e., ${\cal F}=\gamma V^{2/3}a(\xi)+\varepsilon E^2Vv(\xi)$. Here, the first and second terms are associated with surface energy and electrical energy, respectively, wherein $a(\xi)$ and $v(\xi)$ are non-dimensional shape functions. Each steady-state droplet shape corresponds to a minimum of the free-energy, ${\cal F}$, such that $d{\cal F}=0$. Denoting the critically stable droplet shape by $\xi=\xi_0$, $d{\cal F}=0$ implies $\gamma V_{cr}^{2/3}a'(\xi_0)+\varepsilon E^2_{cr}V_{cr}v'(\xi_0)=0$, which in turn yields $\varepsilon E_{cr}^2V_{cr}^{1/3}/\gamma =-a'(\xi_0)/v'(\xi_0)$. As the variation is arbitrary, $a'(\xi_0)/v'(\xi_0)$ is a constant, albeit electric field dependent. From our simulations, we found that $a'(\xi_0)/v'(\xi_0) \approx (E_{cr}^*)^{1.67}$, where $E_{cr}^*=E_{cr}/\sqrt{\gamma/\varepsilon H_0}$. Thus, the final scaling law is obtained as $E_{cr}^*\sim H_0^3V_{cr}^{-1}$. The scaling law for pinned droplets with a fixed wetting radius under electric field was found to have an exponent of $-2$ \cite{beroz2019stability}. The deviation of the exponent observed in this study is attributed to the motion of the droplet contact line in response to the electric field. This difference in scaling law exponent can have profound consequences in a variety of applications, including determining the size of raindrops in thunderstorms that generate preferential conduction channels for lightning strikes and power line failure.


To summarise, we have investigated the dynamics of an electrified generic sessile droplet on a lyophilic substrate. The droplet undergoes deformation in the direction of the electric field and becomes unstable once the applied electric field exceeds a critical limit. The critical droplet height, the critical contact angle, the critical droplet radius are all interdependent unknowns in a generic electrowetting scenario. Despite this inherent complexity in the underlying physics of generic electrified sessile droplets, it is remarkable that a universal scaling law characterizes their stability limit.  While \citet{beroz2019stability} cleverly introduced a scaling law for the case of a pinned droplet with a fixed wetting radius, our current approach extends their analysis of the stability limit of electrified droplets for a more generic case, wherein the droplet radius is free to move in response to the imposed field. The results of the current study may lead to more fruitful research endeavors in many areas involving electrified droplets. For instance, in addition to the several important applications discussed earlier, the results of the current study can also be used as an effective anti-icing mechanism on hydrophilic surfaces, which is a challenging task. A natural extension of the current study is to consider scenarios with droplets that contain free charge carriers. Most technological applications and natural phenomena have droplets with dissolved ionic species, and this work is expected to lay the groundwork for research in that direction.\\

D.S.P. and K.C.S. acknowledge financial support from Science and Engineering Research Board, India through grants SRG/2020/00242 and CRG/2020/000507, respectively. We also thank the reviewers for their valuable suggestions that helped to improve the quality and readability of the manuscript.


%

\begin{center}
{\Large \bf Supplementary information}
\end{center}

\section{Governing equations}
The electrohydrodynamics of a sessile droplet is governed by the continuity and the Navier-Stokes equations, which are given by
\begin{eqnarray}
\nabla\cdot\vec{v}=0,~~\mbox{and}  \qquad\\
\rho\left(\frac{\partial\vec{v}}{\partial t}+\vec{v}\cdot\nabla\vec{v}\right)=\nabla\cdot\textbf{T}
\label{eq:govfluid1}
\end{eqnarray}
Here, $\vec{v} = (u,v,w)$ is the velocity vector, wherein $u$, $v$ and $w$ are the components in the $x$, $y$ and $z$ directions, respectively and $t$ is time. The Cauchy stress tensor is given by $\textbf{T}=-p\textbf{I}+\mu(\nabla\vec{v}+\nabla\vec{v}^{T})$, where $p$ denotes the isotropic pressure field and $\textbf{I}$ is the identity tensor. The droplet is assumed to be a perfect conductor, and therefore is at uniformly constant potential, $\Phi_0$, as imposed at the bottom electrode. The electric field in the ambient medium, $\vec{E}_2=-\nabla\Phi_2$ satisfies Gauss's law, which for a charge-free domain reduces to:
\begin{equation}\label{eq:field}
\nabla^{2}\Phi_2=0,
\end{equation}
subjected to the boundary conditions given by $\Phi_2=\Phi_0$ at $z=h(x,y,t)$ (i.e., at the droplet interface) and $\Phi_2=0$ at $z=H_0$ (i.e., at the top electrode).

The interface speed ($\mathcal{U}$) and the unit normal vector ($\vec{n}$) are given by
\begin{subequations}
\begin{eqnarray}
\mathcal{U}&=&\frac{\frac{\partial h}{\partial t}}{\left[1+\left(\frac{\partial h}{\partial x}\right)^2+\left(\frac{\partial h}{\partial y}\right)^2\right]^{1/2}}, ~ \text{and} \nonumber \\ 
 \vec{n}&=&\frac{-\frac{\partial h}{\partial x}\hat{i}_x-\frac{\partial h}{\partial y}\hat{i}_y+\hat{i}_z}{\left[1+\left(\frac{\partial h}{\partial x}\right)^2+\left(\frac{\partial h}{\partial y}\right)^2\right]^{1/2}},
\end{eqnarray}
respectively. The interfacial tangent vectors ($\vec{t}_x$) and ($\vec{t}_y$) are given by
\begin{eqnarray}
\vec{t}_x&=&\frac{\hat{i}_x+\frac{\partial h}{\partial x}\hat{i}_z}{\left[1+\left(\frac{\partial h}{\partial x}\right)^2+\left(\frac{\partial h}{\partial y}\right)^2\right]^{1/2}}, ~ \text{and}  \nonumber \\     
\vec{t}_y&=&\frac{\hat{i}_y+\frac{\partial h}{\partial y}\hat{i}_z}{\left[1+\left(\frac{\partial h}{\partial x}\right)^2+\left(\frac{\partial h}{\partial y}\right)^2\right]^{1/2}},
\end{eqnarray}
\end{subequations}
where, $\hat{i}_x$, $\hat{i}_y$ and $\hat{i}_z$ are the unit vectors in the $x$, $y$ and $z$ directions, respectively.

The kinematic condition at the material fluid interface ($z=h$) is given as $\vec{v}\cdot\vec{n}-\mathcal{U}=0$, whence
\begin{equation}
w =\frac{\partial h}{\partial t}+u\frac{\partial h}{\partial x} +v\frac{\partial h}{\partial y}.
\end{equation}
At the interface, the tangential stress balance projected along the $x-z$ and $y-z$ planes and the normal stress balance are given by
\begin{subequations}
\begin{eqnarray}
\vec{n}\cdot\textbf{T}\cdot\vec{t}_x&=0, \label{eq:bctstress1}\\
\vec{n}\cdot\textbf{T}\cdot\vec{t}_y &=& 0, ~{\rm and} \label{eq:bctstress2}\\
\vec{n}\cdot\textbf{T}\cdot\vec{n}-\vec{n}\cdot\textbf{M}_2\cdot\vec{n} &=& -\gamma\nabla_s\cdot\vec{n} \nonumber \\
&+&\frac{2s}{h_f}\left(\frac{h_f^{3}}{h^{3}}-\frac{h_f^{2}}{h^{2}}\right),
\label{eq:bcnstress}
\end{eqnarray} 
\end{subequations}

The Maxwell stress tensor ($\textbf{M}_2$) is the second invariant of the electric field gradient tensor in the ambient medium that takes into account the stresses arising due to the applied electric field. This is given by 
\begin{equation}
\textbf{M}_2=\varepsilon \left[\vec{E}_2\vec{E}_2-{1 \over 2} (\vec{E}_2\cdot\vec{E}_2)\textbf{I}\right],
\end{equation}
where $\vec{E}_2$ is the electric field in droplet phase, $\varepsilon$ is the permittivity of the ambient medium.

As evident from Eq. (\ref{eq:bcnstress}), a jump in the normal component of the Maxwell stress between the two fluids at the interface results in additional stress. The last term in Eq. (\ref{eq:bcnstress}) is the conjoining-disjoining potential, with $s=\gamma\left(1-\cos \theta\right)$ being the wetting parameter that determines the equilibrium static contact angle $(\theta)$ of the droplet \cite{Gomba2010}. The thickness that minimises the conjoining-disjoining potential is $h_f$, which is of the order of magnitude of the height of thin precursor film.

In the thin-film approximation, the characteristic length scale in the $x$ and $y$ directions is much larger than that in the $z$ direction, which is taken to be $V^{1/3}$ in our study. Here, $V$ is the volume of the droplet. We introduce $\delta = H_0/V^{1/3}$ for the subsequent simplification. Although the thin film model employed in this study is only expected to be strictly valid for highly wetting droplets, numerical simulations have shown that it can accurately predict droplet dynamics for static contact angle $(\theta)$ up to $65^\circ$ \cite{tembely2019comprehensive}.

The simplified governing equations, by retaining all terms of $\mathcal{O}(\delta)$ and neglecting higher order terms, are given by 
\begin{eqnarray}
   \frac{\partial{u}}{\partial{x}} + \frac{\partial{v}}{\partial{y}} + \frac{\partial{w}}{\partial{z}} &=& 0, \label{continuity}
\\
 \rho \left( \frac{\partial{u}}{\partial{t}}+ u\frac{\partial{u}}{\partial{x}} + v\frac{\partial{u}}{\partial{y}} + w\frac{\partial{u}}{\partial{z}}\right) &=& -\frac{\partial p}{\partial x} +\mu \frac{\partial^2{u}}{\partial{z^{2}}}, \label{xmomentum}
\\
 \rho \left( \frac{\partial{v}}{\partial{t}}+ u\frac{\partial{v}}{\partial{x}} + v\frac{\partial{v}}{\partial{y}} + w\frac{\partial{v}}{\partial{z}}\right) &=& -\frac{\partial p}{\partial y} +\mu\frac{\partial^2{v}}{\partial{z^{2}}}, \label{ymomentum}
\\
 \frac{\partial{p}}{\partial{z}} = 0,   \label{zmomentum} 
\end{eqnarray}
where $\mu$ and $\rho$ are the dynamic viscosity and density of the droplet. The various boundary conditions are
\begin{subequations}
\begin{equation}\label{xstress}
\frac{\partial{u}}{\partial{z}}(h)= 0,~~~~~ \frac{\partial{v}}{\partial{z}}(h)= 0, ~~~~\mbox{and}
\end{equation}
\begin{eqnarray}\label{nstress}
    p(h) &=&-\frac{\varepsilon}{2}\left(\frac{\partial \phi_2}{\partial{z}}\right)^2-\gamma\Bigg[\frac{\partial^2 h}{\partial {x}^2}+\frac{\partial^2 h}{\partial {y}^2}\Bigg] \nonumber \\
    &-&\frac{2\gamma(1-\cos\theta)}{h_f}\Bigg[\frac{h_f^{3}}{h^3}-\frac{h_f^{2}}{h^2}\Bigg].
\end{eqnarray}
\end{subequations}
Here, $\gamma$ is the surface tension. Now integrating the continuity equation (\ref{continuity}) and using the kinematic condition, we obtain the evolution equation for the interface position ($h$) as 
\begin{equation}
\frac{\partial q_x}{\partial x}+\frac{\partial q_y}{\partial y}+\frac{\partial h}{\partial t}=0,
\end{equation}
where the flow rate in the $x$ and $y$ directions are given by
\begin{equation}
q_x= \int_0^h u dz
\mbox{ and }
q_y= \int_0^h v dz, ~{\rm respectively.}
\end{equation}
Integrating Eqs. (\ref{xmomentum})-(\ref{zmomentum}) and eliminating pressure, $p$, we obtain
\begin{subequations}
\begin{eqnarray}\label{redx}
     \rho \Bigg( \frac{\partial{{u}}}{\partial{t}}&+& {u}\frac{\partial{{u}}}{\partial{x}} + {v}\frac{\partial{{u}}}{\partial{y}} + {w}\frac{\partial{{u}}}{\partial{z}}\Bigg) = -\frac{\varepsilon}{2}\frac{\partial}{\partial x}\left(\frac{\partial \phi_2}{\partial{z}}\right)^2  \nonumber \\ &+& \gamma\frac{\partial}{\partial x}\Bigg[\frac{\partial^2 h}{\partial {x}^2}+\frac{\partial^2 h}{\partial {y}^2}\Bigg]
      \nonumber \\ &+& \frac{2\gamma(1-\cos\theta)}{h_f} \frac{\partial}{\partial x}\Bigg(\frac{h_f^{3}}{h^3}-\frac{h_f^{2}}{h^2}\Bigg)+\mu\frac{\partial ^2{u}}{\partial z^2},
 \end{eqnarray}
\begin{eqnarray}\label{redy}
     \rho \Bigg( \frac{\partial{{v}}}{\partial{t}} &+& {u}\frac{\partial{{v}}}{\partial{x}} + {v}\frac{\partial{{v}}}{\partial{y}} + {w}\frac{\partial{{v}}}{\partial{z}}\Bigg) = -\frac{\varepsilon}{2}\frac{\partial}{\partial y}\left(\frac{\partial \phi_2}{\partial{z}}\right)^2 \nonumber \\ &+& \gamma\frac{\partial}{\partial y}\Bigg[\frac{\partial^2 h}{\partial {x}^2}+\frac{\partial^2 h}{\partial {y}^2}\Bigg] \nonumber \\ &+& \frac{2\gamma(1-\cos\theta)}{h_f} \frac{\partial}{\partial y}\Bigg(\frac{h_f^{3}}{h^3}-\frac{h_f^{2}}{h^3}\Bigg)+\mu \frac{\partial ^2{v}}{\partial z^2},
 \end{eqnarray}
 \end{subequations}
 
\section{WRIBL Model}

We decompose the velocity components $u$ and $v$ as follows:
\begin{subequations}
\begin{eqnarray}
    u(x,y,z,t)&=&\hat{u}(x,y,z,t)+\Tilde{u}(x,y,z,t), \\
    v(x,y,z,t)&=&\hat{v}(x,y,z,t)+\Tilde{v}(x,y,z,t),
\end{eqnarray}
\end{subequations}
where, $\hat{u}$ and $\hat{v}$ are the $\mathcal{O}(1)$ contributions and $\tilde{u}$ and $\tilde{v}$ denote their corresponding $\mathcal{O}(\delta)$ corrections. The leading order velocities are chosen to be locally parabolic at their respective coordinate position. This assumption is valid up to moderate Reynolds numbers. The following equations are used to find $\hat{u}$ and $\hat{v}$ as,
\begin{subequations}
\begin{eqnarray}
    \frac{\partial ^2\hat{u} }{\partial z ^2} &=& K_u ,~\hat{u}\mid_0=0,\frac{\partial \hat{u} }{\partial z }\mid_h=0 , ~ q_x= \int_0^h \hat{u} dz, \label{uhat}
\\
    \frac{\partial ^2\hat{v} }{\partial z ^2} &=& K_v ,~\hat{v}\mid_0=0,\frac{\partial \hat{v} }{\partial z }\mid_h=0, ~ q_y= \int_0^h \hat{v} dz. \label{vhat}
\end{eqnarray}
\end{subequations}
The terms $K_u$ and $K_v$  are introduced so that the leading order velocities proﬁle is locally parabolic, and is obtained in terms of the flow rates, $q_x$ and $q_y$ using the integral constraint in Eqs.(\ref{uhat}) and (\ref{vhat}). Note that $\int_0^h \tilde{u} dz=0$ and $\int_0^h \tilde{v} dz=0$. From Eq. (\ref{continuity}),
\begin{equation}\label{what}
    \hat{w}=-\int_0^z \frac{\partial \hat{u}}{\partial x} dz-\int_0^z \frac{\partial \hat{v}}{\partial y} dz.
\end{equation}
Next, we substitute the expressions for velocities obtained from Eqs. (\ref{uhat}), (\ref{vhat}) and (\ref{what}) in Eqs. (\ref{redx}) and (\ref{redy}), and neglect all terms of $\mathcal{O}(\delta \tilde{u})= \mathcal{O}(\delta^2)$, $\mathcal{O}(\delta \tilde{v})=\mathcal{O}(\delta^2)$ or smaller. We take integral with a suitable weight function using the Galerkin method. The weight function, $F$ can be deﬁned as:
\begin{equation}\label{wf}
    \frac{\partial ^2 F }{\partial z ^2}= 1, ~ F\mid_0=\,0, ~ \frac{\partial F }{\partial z }\mid_h=\,0.
\end{equation}
From the above, we obtain the weight function, $F={z^2/2}-zh(x,t)$.
After taking the weighted integral of diffusion terms in Eqs. (\ref{redx}) and (\ref{redy}), and using the weight function given in Eq. (\ref{wf}), we get
\begin{subequations}
\begin{equation}\label{uf}
    \int_0^h F \frac{\partial ^2\hat{u}}{\partial z^2} dz = F \frac{\partial \hat{u} }{\partial z }\mid_h+q_x,
\end{equation}
\begin{equation}\label{vf}
    \int_0^h F \frac{\partial ^2\hat{v}}{\partial z^2} dz = F \frac{\partial \hat{v} }{\partial z }\mid_h+q_y.
\end{equation}
\end{subequations}
The expression for $\frac{\partial \hat{u} }{\partial z },\frac{\partial \hat{v} }{\partial z } $ in Eqs. (\ref{uf}) and (\ref{vf}) can be obtained using Eq. (\ref{xstress}). The final set of evolution equations for the interface position ($h$), the depth integrated flow rate in the $x$ direction ($q_x$) and $y$ direction ($q_y$) are given by
\begin{subequations}\label{alleq}
\begin{equation}
\frac{\partial h}{\partial t}+\frac{\partial q_x}{\partial x}+\frac{\partial q_y}{\partial y}=0, 
\end{equation}
\begin{eqnarray}
 \int_0^h &\rho& F\Bigg( \frac{\partial {\hat u}}{\partial t}+ {\hat u} \frac{\partial {\hat u}}{\partial x} + {\hat v} \frac{\partial {\hat u}}{\partial y} + {\hat w} \frac{\partial {\hat u}}{\partial z} \Bigg)dz = \mu q_x \nonumber \\ &+&\Bigg[\gamma\frac{\partial}{\partial x}\Bigg(\frac{\partial^2 h}{\partial {x}^2}+\frac{\partial^2 h}{\partial {y}^2}\Bigg) -\frac{\varepsilon}{2}\frac{\partial}{\partial x}\left(\frac{\partial \phi_2}{\partial{z}}\right)^2\newblock \nonumber \\ &+&\frac{2\gamma(1-\cos\theta)}{h_f} \frac{\partial}{\partial x}\Bigg(\frac{h_f^{3}}{h^3}-\frac{h_f^{2}}{h^2}\Bigg)\Bigg]\int_0^h F dz,
\end{eqnarray}
\begin{eqnarray}
    \int_0^h &\rho& F\Bigg( \frac{\partial {\hat v}}{\partial t}+ {\hat u} \frac{\partial {\hat v}}{\partial x} + {\hat v}\frac{\partial {\hat v}}{\partial y} + \hat{w}\frac{\partial{\hat{v}}}{\partial{z}}\Bigg)dz =  \mu q_y 
    \nonumber \\
    &+&\Bigg[\gamma\frac{\partial}{\partial y}\Bigg(\frac{\partial^2 h}{\partial {x}^2}+\frac{\partial^2 h}{\partial {y}^2}\Bigg) -\frac{\varepsilon}{2}\frac{\partial}{\partial y}\left(\frac{\partial \phi_2}{\partial{z}}\right)^2\newblock  \nonumber \\ &+&\frac{2\gamma(1-\cos\theta)}{h_f} \frac{\partial}{\partial y}\Bigg(\frac{h_f^{3}}{h^3}-\frac{h_f^{2}}{h^2}\Bigg)\Bigg]\int_0^h F dz.
\end{eqnarray}
\end{subequations}
The length of the computational domain, $L$, in the $x$ and $y$ directions are assumed to be the same, which is set to $2L=8$, with the droplet's center at $(0,0,0)$. 

The governing evolution equations (\ref{alleq}) are solved using the following initial conditions:
\begin{equation}
h(x,y,0)=h_e{}_q (x,y),~~ q_x(x,y,0)=0, \mbox{ and}~~ q_y(x,y,0)=0,
\end{equation}
where $h_e{}_q (x,y)$ is the equilibrium shape of the sessile droplet in the absence of an electric field, which is obtained by performing a transient simulation in the absence of electric field until a steady state droplet configuration is achieved. The initial conditions and spatially periodic boundary conditions used to obtain $h_e{}_q (x,y)$ are given by:
\begin{eqnarray}
 q_{x,in}(x,y)=0,~ q_{y,in}(x,y)=0, \nonumber \\ 
 h_{in}(x,y)=\begin{cases}
	(1-x^2-y^2)+h_f,& \text{if} |x^2+y^2|\leq 1,\\
	h_f,& \text{if} |x^2+y^2|>1.
	\end{cases}
\end{eqnarray}
\begin{subequations}\label{bc}
\begin{eqnarray}
q_x(-L,y)=q_x(L,y), ~~ q_y(-L,y)=q_y(L,y), \nonumber \\  h(=-L,y)=h(L,y),~~ \frac{\partial^n h}{\partial x^n}(-L,y)=  \nonumber \\  \frac{\partial^n h}{\partial x^n}(L,y), ~n=1,2.
\end{eqnarray}
\begin{eqnarray}
  q_x(x,-L)=q_x(x,L), ~~q_y(x,-L)=q_y(x,L), \nonumber \\ h(x,-L)=h(x,L), \nonumber \\ \frac{\partial^n h}{\partial y^n}(x,-L)= \frac{\partial^n h}{\partial y^n}(x,L),~n=1,2.
\end{eqnarray}
\end{subequations}
The initial condition ensures that the droplet volume remains constant for droplets of different contact angle. The Fourier spectral collocation technique is used to assure high-order spatial resolution and that the periodic spacial boundary conditions are satisfied. The NDSolve, a Mathematica v.12.0 function, is used to perform time integration using adaptive time stepping. The 3D simulations are performed for three choices of grid points ($N_x\times N_y$) in the Fourier spectral collocation technique, viz., $(50\times50)$, $(60\times60)$, and $(70\times60)$. As the mean error in droplet deformation between $(60\times60)$ and $(70\times60)$ was found to be less than $1\%$, we choose $(60\times60)$ grids for all 3D simulations.

\end{document}